\documentclass[aps,prb,twocolumn,amsmath,amssymb,superscriptaddress]{revtex4-2}
\usepackage{graphicx}
\usepackage{amssymb}
\usepackage{color}
\usepackage{amsmath}
\usepackage{float}
\usepackage{gensymb}

\usepackage{epstopdf}
\usepackage{hyperref}

\newcommand{\beq}{\begin{eqnarray}}
	\newcommand{\eeq}{\end{eqnarray}}

\begin{document}
	\title{Winding feature and thermal evolution of the Dirac magnons in CrI$_3$
	}
	\author{Weiliang~Yao}
	\email{weiliangyao@outlook.com}
	\affiliation{Department of Physics and Astronomy, Rice University, Houston, Texas 77005, USA}
	\affiliation{Rice Laboratory for Emergent Magnetic Materials and Smalley-Curl Institute, Rice University, Houston, Texas 77005, USA}
	\author{Matthew~B.~Stone}
	\affiliation{Neutron Scattering Division, Oak Ridge National Laboratory, Oak Ridge, Tennessee 37831, USA}
	\author{Colin~L.~Sarkis}
	\affiliation{Neutron Scattering Division, Oak Ridge National Laboratory, Oak Ridge, Tennessee 37831, USA}
	\author{Yi~Li}
	\affiliation{Center for Advanced Quantum Studies and Department of Physics, Beijing Normal University, Beijing 100875, People’s Republic of China}
	\author{Ruixian~Liu}
	\affiliation{Center for Advanced Quantum Studies and Department of Physics, Beijing Normal University, Beijing 100875, People’s Republic of China}
	\author{Xingye~Lu}
	\affiliation{Center for Advanced Quantum Studies and Department of Physics, Beijing Normal University, Beijing 100875, People’s Republic of China}
	\author{Pengcheng~Dai}
	\email{pdai@rice.edu}
	\affiliation{Department of Physics and Astronomy, Rice University, Houston, Texas 77005, USA}
	\affiliation{Rice Laboratory for Emergent Magnetic Materials and Smalley-Curl Institute, Rice University, Houston, Texas 77005, USA}
	\date{\today}
	
	\begin{abstract}
		Two-dimensional honeycomb lattice ferromagnet chromium tri-iodide (CrI$_3$) has attracted tremendous interest because it retains ferromagnetism down to the monolayer limit and hosts intriguing topological magnons. As a prototypical van der Waals magnet, CrI$_3$ provides an ideal platform for exploring the interplay between reduced dimensionality, magnetic order, and nontrivial spin excitations. Here, using inelastic neutron scattering together with improved sample quality, we uncover the magnon winding feature around the $K$-point of the hexagonal Brillouin zone, a key signature of Dirac magnons. In addition, we find that the magnon energy follows a $T^2$-renormalization behavior at elevated temperatures, consistent with magnon-magnon interactions. These results provide previously missing information on the magnon spectrum of CrI$_3$ and further consolidate the topological nature of its spin excitations.
	\end{abstract}
	
	\maketitle
	\section{Introduction}
\begin{figure*}[t!]
		\centering{\includegraphics[width=0.95\textwidth]{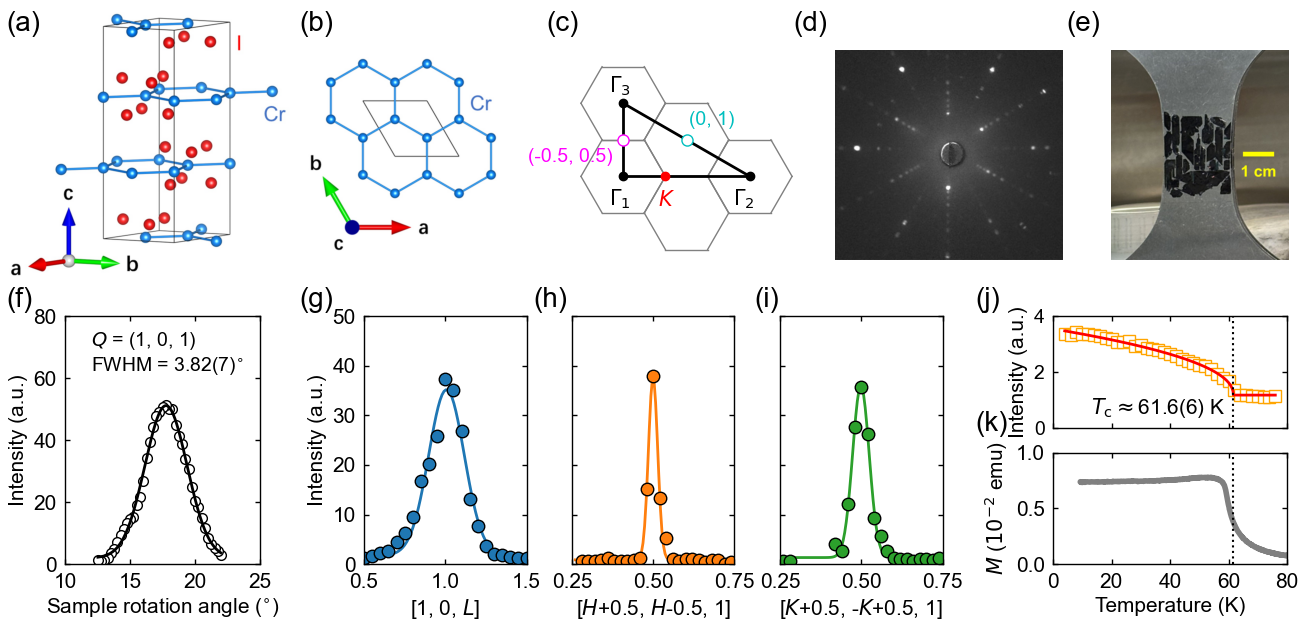}}
		\caption{Structural and magnetic characterization of CrI$_3$. (a) Crystal structure of CrI$_3$ with Cr (blue) and I (red) atoms. (b) Honeycomb network of Cr ions in the $ab$-plane. (c) Schematic of the hexagonal Brillouin zones indicating the momentum-space path used for the cuts in Fig. \ref{fig3}. (d) Laue diffraction pattern of a representative single crystal. (e) Photograph of the coaligned single-crystal assembly used for the measurements. (f) Rocking curve of the (1, 0, 1) Bragg peak. (g)-(i) Intensities of the (1, 0, 1) Bragg peak along [0, 0, $L$], [$H$, $H$, 0], and [$K$, $-K$, 0] directions, respectively. The solid curves in (f)-(i) are the fits with a Gaussian profile. (j) Temperature dependence of the Bragg peak intensity at (1, 0, 1). The solid curve is the fit to the data as described in the text. (k) Temperature dependence of the magnetization under a magnetic field applied along the $c$-axis of the sample. The vertical dotted lines in (j) and (k) indicate $T_{\rm{c}}$.}
		\label{fig1}
	\end{figure*}

	The discovery of long-range ferromagnetic order in atomically thin van der Waals materials has opened a new avenue for exploring collective phenomena in reduced dimensions \cite{BurchNature2018,GibertiniNN2019,doi:10.1021/acsnano.1c09150,ParkRMP2026}. Among these, CrI$_3$ [Fig. \ref{fig1} (a)] has emerged as a model system because it exhibits robust ferromagnetic order down to the monolayer limit \cite{HuangNature2017}, 
	and provides a clean platform for studying the interplay between low dimensionality, anisotropy, and spin excitations \cite{McGuireCM2015,Lado2Dmater2017}. Its honeycomb arrangement of Cr ions [Fig. \ref{fig1} (b)] further makes CrI$_3$ particularly attractive for investigating bosonic band topology, since a honeycomb lattice naturally supports symmetry-protected band crossings and associated topological magnon phenomena \cite{OwerreJPCM2016,PershogubaPRX2018,MookPRX2021,annurev:/content/journals/10.1146/annurev-conmatphys-031620-104715}. Both of these attributes make the system potentially useful for dissipationless spintronics.
	
	A major prediction for ferromagnetic honeycomb magnets is the existence of Dirac magnons near the $K$-points of the hexagonal Brillouin zone \cite{OwerreJPCM2016,FranssonPRB2016,annurev:/content/journals/10.1146/annurev-conmatphys-031620-104715}. In close analogy to Dirac electrons in graphene \cite{NetoRMP2009}, these excitations arise from the two-sublattice structure of the honeycomb lattice and exhibit a linear dispersion relation between energy and momentum near the Dirac point. In particular, the magnon eigenvectors are expected to generate a winding of the spectral weight around the $K$-point \cite{ShivamArxiv2017,annurev:/content/journals/10.1146/annurev-conmatphys-031620-104715}, which provides a direct experimental fingerprint of the topological nature of the excitations. Although CrI$_3$ has been extensively studied by inelastic neutron scattering (INS) \cite{ChenPRX2018,ChenPRB2020,ChenPRX2021,PhysRevB.109.024409,PhysRevB.110.144439}, which is a direct experimental probe for the magnon spectrum, this winding feature has remained difficult to resolve experimentally, in part because it requires both high-quality single-crystal samples and sufficient momentum-space resolution.
	
	Another important issue concerns the thermal evolution of the magnon spectrum in CrI$_3$ \cite{PershogubaPRX2018,EtoArxiv2025}. Because CrI$_3$ is a quasi-two-dimensional ferromagnet with relatively low magnetic energy scales, thermal fluctuations can strongly renormalize the magnon spectrum on approaching the ordering temperature. Understanding how the magnon energies soften and how the excitations broaden with increasing temperature is essential not only for a microscopic description of CrI$_3$, but also for clarifying the role of magnon-magnon interactions in topological magnetic systems \cite{LuPRL2021}.
	
	In this work, we use INS and coaligned single crystals of CrI$_3$ with improved sample mosaic to address these questions. We directly resolve the winding of the magnon intensity around the $K$-point of the Brillouin zone and track the thermal renormalization of the magnon spectrum over a broad temperature range. Our results establish a key missing experimental signature of the Dirac magnons in CrI$_3$ and show that the magnon energies exhibit an approximately $T^2$-renormalization at elevated temperatures, consistent with interaction effects among thermally excited magnons.
	
	\section{Experimental Methods}
	
	High-quality CrI$_3$ single crystals were grown using a chemical vapor transport method slightly modified from previous reports \cite{McGuireCM2015,HuangNature2017}. In a typical growth, a mixture of 0.385 gram chromium powder and 3.10 gram iodine was loaded into a quartz tube (outer diameter 35 mm, wall thickness 1.5 mm), which was then evacuated and sealed under vacuum. The sealed portion of the tube was approximately 22 cm in length. The tube was placed in a one-zone furnace with the starting materials positioned at the hot end. Since the central region of the furnace has nearly uniform temperature, the tube was shifted 8 cm away from the center to create a temperature gradient. The furnace was programmed to ramp to 630 $^{\circ}$C within 7 hours, be held at this temperature for 96 hours, and then be switched off to cool to room temperature. CrI$_3$ crystals can be harvested at the cold end of the tube. The obtained single crystals were characterized with X-ray Laue diffraction and magnetization measurements.
	
	The INS experiment was conducted at the Fine-Resolution Fermi Chopper Spectrometer (SEQUOIA) installed at the Spallation Neutron Source (SNS) of Oak Ridge National Laboratory \cite{GranrothPhysicaB2006,GranrothJPCS2010}. About 0.7 gram of CrI$_3$ single crystals were coaligned with the crystallographic ($H$, $H$, $L$) being put in the horizontal scattering plane. These samples were mounted onto a strain cell \cite{LiuNC2025,LiArxiv2025} so that $\sim$1 \% tensile strain was applied about the [$K$, $-K$, 0] direction at low temperatures. Both the samples and the strain device were mounted within aluminum sample cans sealed with an atmosphere of helium gas for thermal exchange. This sample can was mounted to the bottom loading closed cycle refrigerator used at the instrument \cite{StoneSciRep2025}. However, we did not find any differences for the measured spectra with and without the strain applied (see Section III B). We used an incident neutron energy of 30 meV in the high resolution mode for the measurements at 5 K and in the high flux mode for the measurements from 10 K to 85 K. The energy resolutions at zero energy transfer are about 2.3 meV and 0.8 meV for the high-flux mode and high-resolution mode, respectively. During the experiment, we rotated the sample along the [$K$, $-K$, 0] direction over a total range of 70$^{\circ}$ to ensure substantial measurements of excitations throughout the reciprocal space. The counting time was 0.55 C per step for the high-resolution mode measurement and 0.20 C per step for the high-flux mode measurement. These Coulomb values correspond to the charge accumulated on the spallation target. At the facility operating power used for this measurement, 1.8 MW, they correspond to approximately 6.5 min and 3.5 min per step for the high-resolution and high-flux modes, respectively. Data reduction and analysis were carried out using Mantid \cite{Arnold2014} and HORACE \cite{Ewings2016}. Because the exchange interaction between adjacent Cr honeycomb layers is much weaker than the in-plane exchange interaction within the layers \cite{ChenPRX2021}, we integrated the data over the full measured range along [0, 0, $L$] and present the magnon spectrum in the two-dimensional (2D) ($H$, $K$) plane.
	
	\section{Results}
	\subsection{Structural and magnetic characterization}
	
	Fig. \ref{fig1} (d) shows a X-ray Laue diffraction pattern acquired from measurements of a representative single crystal, which confirms six-fold rotational symmetry and the high crystalline quality of the sample. For most single crystals, the in-plane orientation can be identified from their shape: the longer edge is perpendicular to the [$K$, $K$, 0] direction. Fig. \ref{fig1} (e) shows the coaligned crystal assembly used in the INS measurements. The rocking scan performed for the (1, 0, 1) Bragg peak with 30 meV neutrons demonstrates a mosaic spread of only $3.82(7)^\circ$ [Fig. \ref{fig1} (f)], which is much improved from previously used INS samples with mosaic spread over $6^\circ$ \cite{ChenPRX2018,ChenPRX2021}. The intensities of the $(1, 0, 1)$ Bragg peak along three orthogonal directions, [0, 0, $L$], [$H$, $H$, 0], and [$K$, $-K$, 0], are shown in Fig. \ref{fig1}(g)-(i). By fitting the peaks with Gaussian profiles, we obtained the Full-Width-at-Half-Maximums (FWHMs) of 0.26(1) r.l.u., 0.033(1) r.l.u., and 0.060(2) r.l.u. along the [0, 0, $L$], [$H$, $H$, 0], and [$K$, $-K$, 0] directions, respectively, which correspond to 0.082(4) $\rm{\AA}^{-1}$, 0.061(2) $\rm{\AA}^{-1}$, and 0.063(2) $\rm{\AA}^{-1}$, respectively.
	
	The temperature dependence of the intensity of this peak is presented in Fig. \ref{fig1} (j), which shows typical order-parameter behavior. By fitting the data with a power-law function $I = A (T - T_{\rm{c}})^{2\beta} + B$ ($A$ and $B$ are the scale and background constants, respectively, and $\beta$ is the critical exponent of the order parameter), we find $T_{\rm{c}} = 61.6(6)$ K and $\beta = 0.23(1)$ \cite{ChenPRB2020}. The remaining intensity above $T_{\rm{c}}$ reflects the contribution from the structural scattering. This ferromagnetic transition is consistent with the bulk magnetization measurement as shown in Fig. \ref{fig1} (k), which was measured under 0.1 Tesla magnetic field applied along the $c$-axis of the sample.
	
	\subsection{Magnon dispersion and winding feature around the $K$-point}
	
\begin{figure}[t!]
		\hspace{0cm}\includegraphics[width=8.5cm]{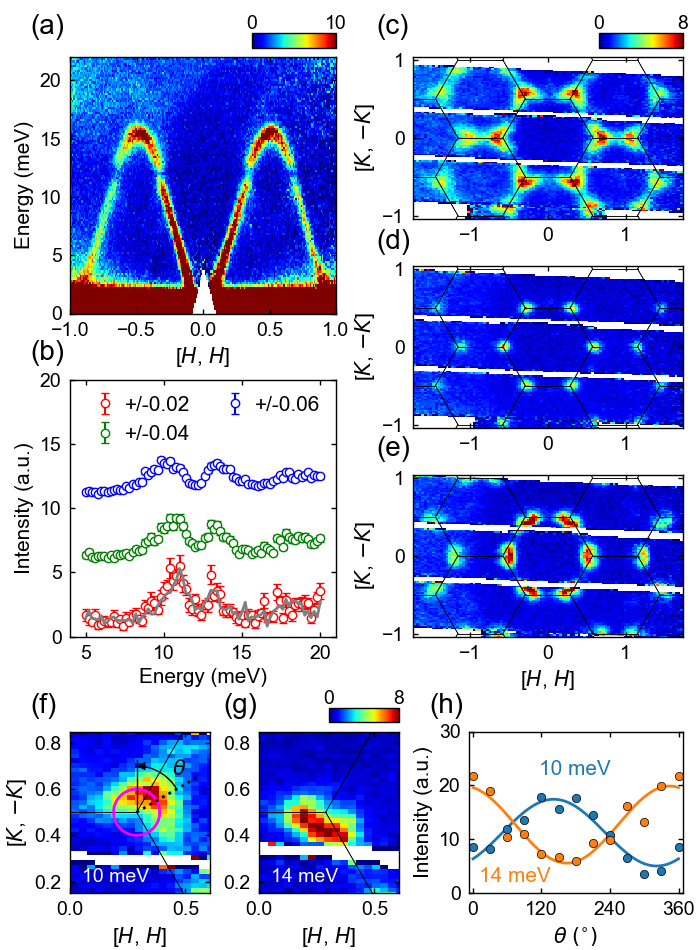}
		\caption{INS-measured magnon spectrum and evidence for the magnon winding feature in CrI$_3$. (a) Neutron-scattering intensity as a function of energy and momentum transfer along [$H$, $H$] direction, displaying the magnon dispersion. (b) Energy dependence of the intensity at the $K$-point (1/3, 1/3) with three data integration ranges along the perpendicular wave-vectors shown in the legend in units of reciprocal lattice units (r.l.u.). The grey curve is the data taken without in-plane strain applied on the sample (see the text) for 0.02 r.l.u. integration. (c)-(e) Constant-energy maps in the plane at 14 meV, 12 meV, and 10 meV, respectively. The black hexagons denote Brillouin-zone boundaries. (f) and (g) Intensity maps near $K$-point at 10 meV and 14 meV, respectively, integrated along the [0, 0, $L$] directio from $-10$ to 10 r.l.u.. (h) Intensity as a function of winding angle ($\theta$) extracted from the data in (f) and (g), where the winding angle is defined in panel (f). The phase shift between the 10- and 14-meV profiles demonstrates the winding of the magnon intensity around the $K$-point [magenta circle in (f)]. The error bars of these data points are smaller than the symbol size. Solid lines are fits to a simple phase shifted cosine function plus a constant value.}
		\label{fig2}
	\end{figure}

	Fig. \ref{fig2} (a) presents the INS-measured spin wave spectrum along the $[H, H]$ direction at 5 K, which is much lower than the $T_{\rm{c}}$. These data were acquired with $E_{\rm{i}}$ = 30 meV neutrons. At the $K$-point of the Brillouin zone ($H = \pm \frac{1}{3}$ and $\pm\frac{2}{3}$), we can clearly resolve the magnon Dirac gap around 12 meV, which is consistent with previous reports \cite{ChenPRX2018,ChenPRX2021}. We note that the presence or absence of a Dirac gap in similar systems CrCl$_3$ and CrBr$_3$ has been actively discussed \cite{CaiPRB2021,NikitinPRL2022,DoPRB2022}. An important caveat is that this conclusion may depend on the integration width used perpendicular to the momentum trajectory \cite{DoPRB2022}. To test whether the observed magnon Dirac gap in CrI$_3$ is affected by this extrinsic effect, we present the energy dependence of the intensity around the gap with three different integration ranges in Fig. \ref{fig2} (b). We can see that the energy gap is always $\sim$2 meV, regardless of the integration range. This further confirms the existence of a magnon Dirac gap in CrI$_3$. In addition, by comparing with the data without the in-plane strain [the grey curve in Fig. \ref{fig2} (b)], we confirm that $\sim$1 \% strain does not affect the observed Dirac gap.
	
	Having confirmed the Dirac gap of the magnon spectrum, we turn to its structure in reciprocal space. Fig. \ref{fig2} (c) shows the constant energy slice around 14 meV, where we can see that the intensity is concentrated around the $K$-points outside of the first Brillouin zone. The intensity center moves to almost the $K$-points and it becomes much weaker at 12 meV, roughly at the Dirac gap [Fig. \ref{fig2} (d)]. Moving to lower energy at 10 meV [Fig. \ref{fig2} (e)], the intensity again shifts away from the $K$-points but at the inside of the first Brillouin zone. Such observation suggests that the intensity distribution at 12 meV and 10 meV (above and below the Dirac gap, respectively) is almost opposite when winding around the $K$-point. Views of the intensity at an equivalent $K$-point $(-1/3, 2/3)$ are presented in Fig. \ref{fig2} (f) and (g). This winding feature is regarded as a key character for topological magnons, which has also been observed in CoTiO$_3$ \cite{ElliotNC2021}, CrBr$_3$ \cite{NikitinPRL2022}, and elemental gadolinium \cite{ScheiePRL2022,ScheiePRB2022}. Note that the existence of this feature does not depend on whether there is a gap in the spectrum at this wave-vector.

	\begin{figure}[t!]
		\hspace{0cm}\includegraphics[width=8.5cm]{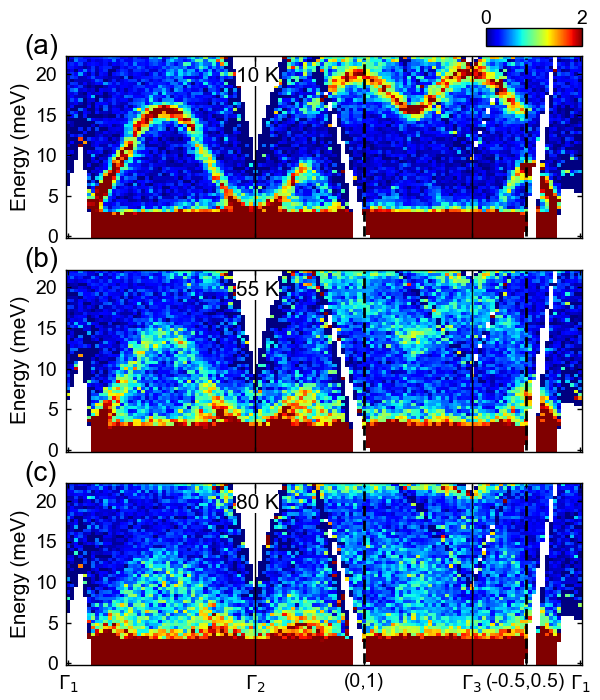}
		\caption{INS spectra of CrI$_3$ along equivalent $\Gamma$-point directions at different temperatures. (a)-(c) Energy-momentum slices measured along the trajectory $\Gamma_1-\Gamma_2-\Gamma_3-\Gamma_1$ [see Fig. \ref{fig1} (c)] at 10 K, 55 K, and 80 K, respectively.}
		\label{fig3}
	\end{figure}
	
	\begin{figure}[t!]
		\hspace{0cm}\includegraphics[width=8.7cm]{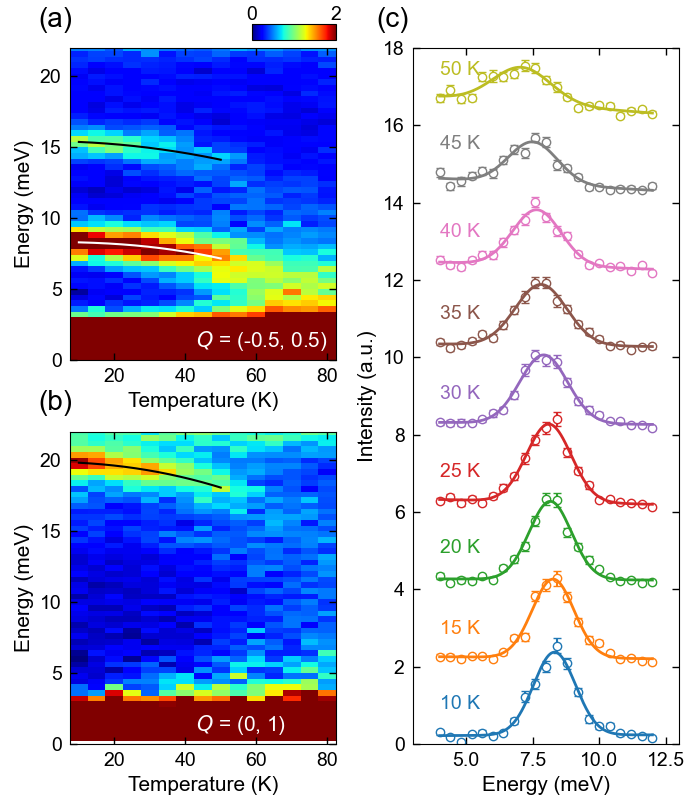}
		\caption{Thermal renormalization of the magnon spectrum of CrI$_3$. (a) and (b) INS intensity as a function of temperature and energy at $\mathbf{Q} = (-0.5, 0.5)$ and $(0, 1)$, respectively. Solid curves are the fits to the extracted energy with power-law function with an offset as described in the text. (c) Representative energy dependence of the intensities at $\mathbf{Q} = (-0.5, 0.5)$ measured at selected temperatures. The data are vertically offset for clarity; open circles denote the measured intensities and solid lines are fits with a Gaussian profile.}
		\label{fig4}
	\end{figure}
	
	To quantify the effect, we extract the intensity as a function of winding angle around the $K$-point [magenta circle in Fig. \ref{fig2} (f)]. The angular profiles at 10 meV and 14 meV are approximately described by cosine functions with a clear phase shift [Fig. \ref{fig2} (h)]. This phase shift demonstrates the winding of the magnon spectral weight around the $K$-point and constitutes direct neutron-scattering evidence for the Dirac character of the magnon excitation in CrI$_3$. The observation of this winding feature is important because it probes not only the existence of Dirac magnons, but also the internal phase structure of the magnon eigenstates. In this sense, the neutron intensity pattern provides a direct window into the topological nature of the spin excitation spectrum. The improved sample quality achieved here is crucial for resolving this subtle momentum-dependent signature, which is not available in previous experiments \cite{ChenPRX2018,ChenPRX2021}.
	
	\subsection{Temperature evolution of the magnon spectrum}
	
	To investigate how the magnon excitations evolve with temperature, we measured energy-momentum slices along a trajectory connecting equivalent $\Gamma$ points, as shown in Fig. \ref{fig1} (c). At 10 K, the magnon branches are sharp and well defined across the measured momentum range [Fig. \ref{fig3} (a)]. At 55 K, just below $T_{\rm{c}}$ [Fig. \ref{fig3} (b)], the modes remain visible but become substantially broader, slightly softened, and with reduced spectral contrast. However, the typical quadratic spin-wave dispersion can still be seen close to the zone center. By 80 K [Fig. \ref{fig3} (c)], the excitations are strongly broadened and much less coherent, indicating pronounced damping and loss of well-defined quasiparticle character.
	
	A more quantitative and detailed view of the thermal evolution of the spin-wave modes is provided in Fig. \ref{fig4}. At ${\bf Q}=(-0.5,0.5)$ [see Fig. \ref{fig1} (c) and Fig. \ref{fig3}], both magnon modes at $\sim$8 meV and $\sim$15 meV soften and decrease in intensity on approaching $T_{\rm{c}}$ [Fig. \ref{fig4}(a)]. Similar behavior is observed at ${\bf Q}=(0,1)$ [see Fig. \ref{fig1} (c) and Fig. \ref{fig3}], where only one magnon mode at $\sim$20 meV can be resolved [Fig. \ref{fig4}(b)]. Representative energy dependence of the intensity around $\sim$8 meV at ${\bf Q}=(-0.5,0.5)$ are shown in Fig. \ref{fig4}(c). The peak position moves continuously to lower energy with increasing temperature, and the peak width increases at the same time, reflecting enhanced thermal damping. By fitting the temperature dependence of the energy to a power-law function with an offset $a+bT^{\alpha}$, we find the power-law exponent $\alpha$ to be 2.1(4), 2.4(3), and 2.0(5) for the spin-wave modes at $\sim$8 meV, $\sim$15 meV, and $\sim$20 meV, respectively. This is consistent with $\alpha = 2$ as expected for an interacting spin-wave model \cite{PershogubaPRX2018}. Similar observation has also been made in the study of CrBr$_3$ \cite{NikitinPRL2022}, though no Dirac gap is present in that system.
	
	\section{Discussion}
	
	The winding feature observed around the $K$-point provides direct support for the Dirac-magnon description of CrI$_3$. In a honeycomb ferromagnet, the two magnetic sublattices give rise to magnon eigenvectors with a nontrivial phase relation, and this phase structure is encoded in the neutron-scattering cross section \cite{ShivamArxiv2017,annurev:/content/journals/10.1146/annurev-conmatphys-031620-104715}. The rotation of the intensity pattern between different energies near the zone-boundary magnon is therefore not a trivial matrix-element effect, but a manifestation of the winding of the magnon wave function in momentum space. Our measurements thus supply an experimentally accessible fingerprint of topological magnon band structure beyond the simple observation of dispersive modes.
	
	The thermal evolution of the spectrum reveals a second important aspect of the magnetic dynamics in CrI$_3$. With increasing temperature, the magnons soften, broaden, and eventually become overdamped above $T_{\rm{c}}$. The extracted mode energies show an approximately $T^2$-renormalization at elevated temperatures, consistent with magnon-magnon interaction effects. Physically, as the thermal population of magnons increases, interactions among these bosonic excitations renormalize the spin-wave energies downward and reduces their lifetime. The stronger renormalization of the lower-energy branch suggests that the thermal response depends sensitively on the local dispersion and available phase space for scattering.
	
	These findings have broader implications for the study of topological excitations in van der Waals magnets. First, they show that subtle momentum-dependent signatures of magnon topology can be resolved directly by neutron scattering when sample quality is sufficiently high. Second, they demonstrate that the topological magnon spectrum remains experimentally accessible over a significant temperature range, while also revealing how interaction effects reshape the excitation spectrum upon warming. Since CrI$_3$ is one of the foundational materials in the broader family of 2D magnets, the present results provide an important benchmark for future studies of interaction-driven and temperature-dependent phenomena in magnonic band topology.
	
	More generally, the combination of a honeycomb lattice, strong magnetic anisotropy, and accessible magnetic energy scales makes CrI$_3$ an ideal model system for connecting microscopic neutron-scattering observables with concepts originally developed in topological band theory. Our results therefore help bridge the gap between model expectations for Dirac magnons and experimentally resolved spectral signatures in a real quantum material.\\
	
	\section{Conclusion}
	
	In summary, we have used INS to investigate the momentum and temperature dependence of the magnon spectrum in CrI$_3$. By exploiting improved sample quality, we directly resolve the winding feature of the neutron-scattering intensity around the $K$-point of the Brillouin zone, providing the missing experimental signature of Dirac magnons in this honeycomb ferromagnet. We further show that the magnon spectrum undergoes substantial thermal renormalization, with the mode energies exhibiting an approximately $T^2$-dependence at elevated temperatures, consistent with magnon-magnon interactions. These results fill an important gap in the experimental characterization of CrI$_3$ and further establish it as a prototypical platform for studying topological spin excitations in van der Waals magnets.

	\begin{acknowledgments}
		The neutron scattering and single crystal synthesis work at Rice are supported by US NSF DMR-2401084 and the Robert A. Welch Foundation under Grant No. C-1839, respectively (P.D.).
		A portion of this research used resources at the Spallation Neutron Source, a DOE Office of Science User Facility operated by the Oak Ridge National Laboratory. The beam time was allocated to SEQUOIA on
		Proposal No. IPTS-34835.1.
	\end{acknowledgments}
	
	\bibliographystyle{apsrev4-1}
	\bibliography{CrI3_thermal_reference_2}

@article{BurchNature2018,
	title={Magnetism in two-dimensional van der Waals materials},
	author={Burch, Kenneth S and Mandrus, David and Park, Je-Geun},
	journal={Nature},
	volume={563},
	number={7729},
	pages={47--52},
	year={2018},
	publisher={Nature Publishing Group UK London},
	url={https://doi.org/10.1038/s41586-018-0631-z}
}

@article{GibertiniNN2019,
	title={Magnetic 2D materials and heterostructures},
	author={Gibertini, Magnetic and Koperski, Maciej and Morpurgo, Alberto F and Novoselov, Konstantin S},
	journal={Nature Nanotechnology},
	volume={14},
	number={5},
	pages={408--419},
	year={2019},
	publisher={Nature Publishing Group UK London},
	url={https://doi.org/10.1038/s41565-019-0438-6}
}

@article{doi:10.1021/acsnano.1c09150,
author = {Wang, Qing Hua and Bedoya-Pinto, Amilcar and Blei, Mark and Dismukes, Avalon H. and Hamo, Assaf and Jenkins, Sarah and Koperski, Maciej and Liu, Yu and Sun, Qi-Chao and Telford, Evan J. and Kim, Hyun Ho and Augustin, Mathias and Vool, Uri and Yin, Jia-Xin and Li, Lu Hua and Falin, Alexey and Dean, Cory R. and Casanova, Fèlix and Evans, Richard F. L. and Chshiev, Mairbek and Mishchenko, Artem and Petrovic, Cedomir and He, Rui and Zhao, Liuyan and Tsen, Adam W. and Gerardot, Brian D. and Brotons-Gisbert, Mauro and Guguchia, Zurab and Roy, Xavier and Tongay, Sefaattin and Wang, Ziwei and Hasan, M. Zahid and Wrachtrup, Joerg and Yacoby, Amir and Fert, Albert and Parkin, Stuart and Novoselov, Kostya S. and Dai, Pengcheng and Balicas, Luis and Santos, Elton J. G.},
title = {The Magnetic Genome of Two-Dimensional van der Waals Materials},
journal = {ACS Nano},
volume = {16},
number = {5},
pages = {6960-7079},
year = {2022},
    note ={PMID: 35442017},

URL = { 
    
        https://doi.org/10.1021/acsnano.1c09150
    
    

},
eprint = { 
    
        https://doi.org/10.1021/acsnano.1c09150
    
    

}

}

@article{ParkRMP2026,
	title={2D van der Waals magnets: From fundamental physics to applications},
	ISSN={1539-0756},
	url={http://dx.doi.org/10.1103/2pff-xy6n},
	DOI={10.1103/2pff-xy6n},
	journal={Reviews of Modern Physics},
	publisher={American Physical Society (APS)},
	author={Je-Geun Park and Kaixuan Zhang and Hyeonsik Cheong and Jae Hoon Kim and Carina Belvin and David Hsieh and Honglie Ning and Nuh Gedik},
	year={2026},
	month=feb }

@article{HuangNature2017,
	title={Layer-dependent ferromagnetism in a van der Waals crystal down to the monolayer limit},
	author={Huang, Bevin and Clark, Genevieve and Navarro-Moratalla, Efr{\'e}n and Klein, Dahlia R and Cheng, Ran and Seyler, Kyle L and Zhong, Ding and Schmidgall, Emma and McGuire, Michael A and Cobden, David H and others},
	journal={Nature},
	volume={546},
	number={7657},
	pages={270--273},
	year={2017},
	publisher={Nature Publishing Group UK London},
	url={https://doi.org/10.1038/nature22391}
}

@article{Lado2Dmater2017,
	title={On the origin of magnetic anisotropy in two dimensional CrI3},
	author={Lado, Jose L and Fern{\'a}ndez-Rossier, Joaqu{\'\i}n},
	journal={2D Materials},
	volume={4},
	number={3},
	pages={035002},
	year={2017},
	publisher={IOP Publishing},
	url={https://doi.org/10.1088/2053-1583/aa75ed}
}

@article{McGuireCM2015,
	author = {McGuire, Michael A. and Dixit, Hemant and Cooper, Valentino R. and Sales, Brian C.},
	title = {Coupling of Crystal Structure and Magnetism in the Layered, Ferromagnetic Insulator CrI3},
	journal = {Chemistry of Materials},
	volume = {27},
	number = {2},
	pages = {612-620},
	year = {2015},
	URL = {https://doi.org/10.1021/cm504242t}
}

@article{OwerreJPCM2016,
	url = {https://doi.org/10.1088/0953-8984/28/38/386001},
	year = {2016},
	month = {jul},
	publisher = {IOP Publishing},
	volume = {28},
	number = {38},
	pages = {386001},
	author = {Owerre, S A},
	title = {A first theoretical realization of honeycomb topological magnon insulator},
	journal = {Journal of Physics: Condensed Matter}
}

@article{annurev:/content/journals/10.1146/annurev-conmatphys-031620-104715,
   author = "McClarty, Paul A.",
   title = "Topological Magnons: A Review", 
   journal= "Annual Review of Condensed Matter Physics",
   year = "2022",
   volume = "13",
   number = "Volume 13, 2022",
   pages = "171-190",
   doi = "https://doi.org/10.1146/annurev-conmatphys-031620-104715",
   url = "https://www.annualreviews.org/content/journals/10.1146/annurev-conmatphys-031620-104715",
   publisher = "Annual Reviews",
   issn = "1947-5462",
   type = "Journal Article",
  }

@article{PhysRevB.109.024409,
  title = {Antiferromagnetic-ferromagnetic homostructures with Dirac magnons in the van der Waals magnet ${\mathrm{CrI}}_{3}$},
  author = {Schneeloch, John A. and Daemen, Luke and Louca, Despina},
  journal = {Phys. Rev. B},
  volume = {109},
  issue = {2},
  pages = {024409},
  numpages = {9},
  year = {2024},
  month = {Jan},
  publisher = {American Physical Society},
  url = {https://link.aps.org/doi/10.1103/PhysRevB.109.024409}
}

@article{PhysRevB.110.144439,
  title = {Role of stacking defects on the magnetic behavior of ${\mathrm{CrCl}}_{3}$},
  author = {Schneeloch, John A. and Aczel, Adam A. and Ye, Feng and Louca, Despina},
  journal = {Phys. Rev. B},
  volume = {110},
  issue = {14},
  pages = {144439},
  numpages = {8},
  year = {2024},
  month = {Oct},
  publisher = {American Physical Society},
  url = {https://link.aps.org/doi/10.1103/PhysRevB.110.144439}
}

@article{MookPRX2021,
	title = {Interaction-Stabilized Topological Magnon Insulator in Ferromagnets},
	author = {Mook, Alexander and Plekhanov, Kirill and Klinovaja, Jelena and Loss, Daniel},
	journal = {Phys. Rev. X},
	volume = {11},
	issue = {2},
	pages = {021061},
	numpages = {30},
	year = {2021},
	month = {Jun},
	publisher = {American Physical Society},
	url = {https://link.aps.org/doi/10.1103/PhysRevX.11.021061}
}

@article{PershogubaPRX2018,
	title = {Dirac Magnons in Honeycomb Ferromagnets},
	author = {Pershoguba, Sergey S. and Banerjee, Saikat and Lashley, J. C. and Park, Jihwey and \AA{}gren, Hans and Aeppli, Gabriel and Balatsky, Alexander V.},
	journal = {Phys. Rev. X},
	volume = {8},
	issue = {1},
	pages = {011010},
	numpages = {8},
	year = {2018},
	month = {Jan},
	publisher = {American Physical Society},
	url = {https://link.aps.org/doi/10.1103/PhysRevX.8.011010}
}

@article{LuPRL2021,
	title = {Topological Phase Transitions of Dirac Magnons in Honeycomb Ferromagnets},
	author = {Lu, Yu-Shan and Li, Jian-Lin and Wu, Chien-Te},
	journal = {Phys. Rev. Lett.},
	volume = {127},
	issue = {21},
	pages = {217202},
	numpages = {6},
	year = {2021},
	month = {Nov},
	publisher = {American Physical Society},
	url = {https://link.aps.org/doi/10.1103/PhysRevLett.127.217202}
}

@article{FranssonPRB2016,
	title = {Magnon Dirac materials},
	author = {Fransson, J. and Black-Schaffer, A. M. and Balatsky, A. V.},
	journal = {Phys. Rev. B},
	volume = {94},
	issue = {7},
	pages = {075401},
	numpages = {6},
	year = {2016},
	month = {Aug},
	publisher = {American Physical Society},
	url = {https://link.aps.org/doi/10.1103/PhysRevB.94.075401}
}

@misc{ShivamArxiv2017,
	title={Neutron Scattering Signatures of Magnon Weyl Points}, 
	author={S. Shivam and R. Coldea and R. Moessner and P. McClarty},
	year={2017},
	eprint={1712.08535},
	archivePrefix={arXiv},
	primaryClass={cond-mat.str-el},
	url={https://arxiv.org/abs/1712.08535}, 
}

@article{ChenPRX2018,
	title = {Topological Spin Excitations in Honeycomb Ferromagnet ${\mathrm{CrI}}_{3}$},
	author = {Chen, Lebing and Chung, Jae-Ho and Gao, Bin and Chen, Tong and Stone, Matthew B. and Kolesnikov, Alexander I. and Huang, Qingzhen and Dai, Pengcheng},
	journal = {Phys. Rev. X},
	volume = {8},
	issue = {4},
	pages = {041028},
	numpages = {7},
	year = {2018},
	month = {Nov},
	publisher = {American Physical Society},
	url = {https://link.aps.org/doi/10.1103/PhysRevX.8.041028}
}

@article{ChenPRB2020,
	title = {Magnetic anisotropy in ferromagnetic ${\mathrm{CrI}}_{3}$},
	author = {Chen, Lebing and Chung, Jae-Ho and Chen, Tong and Duan, Chunruo and Schneidewind, Astrid and Radelytskyi, Igor and Voneshen, David J. and Ewings, Russell A. and Stone, Matthew B. and Kolesnikov, Alexander I. and Winn, Barry and Chi, Songxue and Mole, R. A. and Yu, D. H. and Gao, Bin and Dai, Pengcheng},
	journal = {Phys. Rev. B},
	volume = {101},
	issue = {13},
	pages = {134418},
	numpages = {8},
	year = {2020},
	month = {Apr},
	publisher = {American Physical Society},
	url = {https://link.aps.org/doi/10.1103/PhysRevB.101.134418}
}

@article{ChenPRX2021,
	title = {Magnetic Field Effect on Topological Spin Excitations in ${\mathrm{CrI}}_{3}$},
	author = {Chen, Lebing and Chung, Jae-Ho and Stone, Matthew B. and Kolesnikov, Alexander I. and Winn, Barry and Garlea, V. Ovidiu and Abernathy, Douglas L. and Gao, Bin and Augustin, Mathias and Santos, Elton J. G. and Dai, Pengcheng},
	journal = {Phys. Rev. X},
	volume = {11},
	issue = {3},
	pages = {031047},
	numpages = {11},
	year = {2021},
	month = {Aug},
	publisher = {American Physical Society},
	url = {https://link.aps.org/doi/10.1103/PhysRevX.11.031047}
}

@article{Arnold2014,
	title = {{Mantid-Data analysis and visualization package for neutron scattering and $\mu$SR experiments}},
	journal = {Nuclear Instruments and Methods in Physics Research Section A: Accelerators, Spectrometers, Detectors and Associated Equipment},
	volume = {764},
	pages = {156-166},
	year = {2014},
	issn = {0168-9002},
	url = {https://www.sciencedirect.com/science/article/pii/S0168900214008729},
	author = {O. Arnold and others}
}

@article{Ewings2016,
	title = {Horace: Software for the analysis of data from single crystal spectroscopy experiments at time-of-flight neutron instruments},
	journal = {Nuclear Instruments and Methods in Physics Research Section A: Accelerators, Spectrometers, Detectors and Associated Equipment},
	volume = {834},
	pages = {132-142},
	year = {2016},
	issn = {0168-9002},
	url = {https://www.sciencedirect.com/science/article/pii/S016890021630777X},
	author = {R.A. Ewings and A. Buts and M.D. Le and J. {van Duijn} and I. Bustinduy and T.G. Perring}
}

@article{NikitinPRL2022,
	title = {Thermal Evolution of Dirac Magnons in the Honeycomb Ferromagnet ${\mathrm{CrBr}}_{3}$},
	author = {Nikitin, S. E. and F\aa{}k, B. and Kr\"amer, K. W. and Fennell, T. and Normand, B. and L\"auchli, A. M. and R\"uegg, Ch.},
	journal = {Phys. Rev. Lett.},
	volume = {129},
	issue = {12},
	pages = {127201},
	numpages = {6},
	year = {2022},
	month = {Sep},
	publisher = {American Physical Society},
	url = {https://link.aps.org/doi/10.1103/PhysRevLett.129.127201}
}

@article{ScheiePRL2022,
	title = {Dirac Magnons, Nodal Lines, and Nodal Plane in Elemental Gadolinium},
	author = {Scheie, A. and Laurell, Pontus and McClarty, P. A. and Granroth, G. E. and Stone, M. B. and Moessner, R. and Nagler, S. E.},
	journal = {Phys. Rev. Lett.},
	volume = {128},
	issue = {9},
	pages = {097201},
	numpages = {6},
	year = {2022},
	month = {Mar},
	publisher = {American Physical Society},
	url = {https://link.aps.org/doi/10.1103/PhysRevLett.128.097201}
}

@article{ScheiePRB2022,
	title = {Spin-exchange Hamiltonian and topological degeneracies in elemental gadolinium},
	author = {Scheie, A. and Laurell, Pontus and McClarty, P. A. and Granroth, G. E. and Stone, M. B. and Moessner, R. and Nagler, S. E.},
	journal = {Phys. Rev. B},
	volume = {105},
	issue = {10},
	pages = {104402},
	numpages = {12},
	year = {2022},
	month = {Mar},
	publisher = {American Physical Society},
	url = {https://link.aps.org/doi/10.1103/PhysRevB.105.104402}
}

@article{CaiPRB2021,
	title = {Topological magnon insulator spin excitations in the two-dimensional ferromagnet ${\mathrm{CrBr}}_{3}$},
	author = {Cai, Zhengwei and Bao, Song and Gu, Zhao-Long and Gao, Yi-Peng and Ma, Zhen and Shangguan, Yanyan and Si, Wenda and Dong, Zhao-Yang and Wang, Wei and Wu, Yizhang and Lin, Dongjing and Wang, Jinghui and Ran, Kejing and Li, Shichao and Adroja, Devashibhai and Xi, Xiaoxiang and Yu, Shun-Li and Wu, Xiaoshan and Li, Jian-Xin and Wen, Jinsheng},
	journal = {Phys. Rev. B},
	volume = {104},
	issue = {2},
	pages = {L020402},
	numpages = {7},
	year = {2021},
	month = {Jul},
	publisher = {American Physical Society},
	url = {https://link.aps.org/doi/10.1103/PhysRevB.104.L020402}
}

@article{DoPRB2022,
	title = {Gaps in topological magnon spectra: Intrinsic versus extrinsic effects},
	author = {Do, Seung-Hwan and Paddison, Joseph A. M. and Sala, Gabriele and Williams, Travis J. and Kaneko, Koji and Kuwahara, Keitaro and May, Andrew F. and Yan, Jiaqiang and McGuire, Michael A. and Stone, Matthew B. and Lumsden, Mark D. and Christianson, Andrew D.},
	journal = {Phys. Rev. B},
	volume = {106},
	issue = {6},
	pages = {L060408},
	numpages = {6},
	year = {2022},
	month = {Aug},
	publisher = {American Physical Society},
	url = {https://link.aps.org/doi/10.1103/PhysRevB.106.L060408}
}

@article{ElliotNC2021,
	title={Order-by-disorder from bond-dependent exchange and intensity signature of nodal quasiparticles in a honeycomb cobaltate},
	author={Elliot, M and McClarty, Paul A and Prabhakaran, D and Johnson, RD and Walker, HC and Manuel, P and Coldea, R},
	journal={Nature Communications},
	volume={12},
	number={1},
	pages={3936},
	year={2021},
	publisher={Nature Publishing Group UK London},
	url={https://doi.org/10.1038/s41467-021-23851-0}
}

@article{NetoRMP2009,
	title = {The electronic properties of graphene},
	author = {Castro Neto, A. H. and Guinea, F. and Peres, N. M. R. and Novoselov, K. S. and Geim, A. K.},
	journal = {Rev. Mod. Phys.},
	volume = {81},
	issue = {1},
	pages = {109--162},
	numpages = {0},
	year = {2009},
	month = {Jan},
	publisher = {American Physical Society},
	url = {https://link.aps.org/doi/10.1103/RevModPhys.81.109}
}

@article{GranrothPhysicaB2006,
	title = {SEQUOIA: A fine resolution chopper spectrometer at the SNS},
	journal = {Physica B: Condensed Matter},
	volume = {385-386},
	pages = {1104-1106},
	year = {2006},
	url = {https://www.sciencedirect.com/science/article/pii/S0921452606012701},
	author = {G.E. Granroth and D.H. Vandergriff and S.E. Nagler}
}

@inproceedings{GranrothJPCS2010,
	title={SEQUOIA: a newly operating chopper spectrometer at the SNS},
	author={Granroth, GE and Kolesnikov, AI and Sherline, TE and Clancy, JP and Ross, KA and Ruff, JPC and Gaulin, BD and Nagler, SE},
	booktitle={Journal of Physics: Conference Series},
	volume={251},
	number={1},
	pages={012058},
	year={2010},
	url={https://doi.org/10.1088/1742-6596/251/1/012058}
}

@article{LiuNC2025,
	title={Spin correlations in the nematic quantum disordered state of {FeSe}},
	author={Liu, Ruixian and Stone, Matthew B and Gao, Shang and Nakamura, Mitsutaka and Kamazawa, Kazuya and Krajewska, Aleksandra and Walker, Helen C and Cheng, Peng and Yu, Rong and Si, Qimiao and Dai, Pengcheng and Lu, Xingye},
	journal={Nature Communications},
	volume={16},
	pages={5212},
	year={2025},
	publisher={Nature Publishing Group UK London},
	url={https://doi.org/10.1038/s41467-025-60071-2}
}

@article{LiArxiv2025,
	title={Magnetic excitations in biaxial-strain detwinned $\alpha$-RuCl$_{3}$}, 
	author={Yi Li and Yanyan Shangguan and Xinzhe Wang and Ruixian Liu and Chang Liu and Yongqi Han and Zhaosheng Wang and Christian Balz and Ross Stewart and Shun-Li Yu and Jinsheng Wen and Jian-Xin Li and Xingye Lu},
	journal={arXiv preprint arXiv:2509.06753},
	year={2025},
	url={https://arxiv.org/abs/2509.06753}
}

@article{EtoArxiv2025,
	title={Fate of Topological Dirac Magnons in van der Waals Ferromagnets at Finite Temperature},
	author={Eto, Rintaro and Salgado-Linares, Ignacio and Mochizuki, Masahito and Knolle, Johannes and Mook, Alexander},
	journal={arXiv preprint arXiv:2509.13900},
	year={2025},
	url={https://doi.org/10.48550/arXiv.2509.13900}
}

@article{StoneSciRep2025,
	title={Sample changers for direct geometry neutron chopper spectrometers},
	author={Stone, Matthew B and Granroth, Garrett E and Pajerowski, Daniel M and Abernathy, Douglas L and Conner, David L and DeBeer-Schmitt, Lisa and Fanelli, Victor R and Goyette, Richard and Kolesnikov, Alexander I and Mills, Rebecca and others},
	journal={Scientific Reports},
	volume={15},
	number={1},
	pages={31936},
	year={2025},
	publisher={Nature Publishing Group UK London},
	url={https://doi.org/10.1038/s41598-025-17049-3}
}

\end{document}